\documentclass[journal=nalefd,manuscript=letter,layout=twocolumn]{achemso}

\usepackage[version=3]{mhchem} 

\usepackage{graphicx} 
\usepackage{float}
\usepackage{dcolumn}
\usepackage{bm}
\usepackage{array}
\usepackage{amssymb,amsmath}
\usepackage{epstopdf}
\usepackage{xcolor}
\usepackage[dvipsnames]{xcolor}
\usepackage{mathrsfs}
\usepackage[normalem]{ulem}  
\usepackage[bottom]{footmisc}
\usepackage{textcomp, gensymb}
\usepackage{upgreek}

\newcolumntype{C}[1]{>{\centering\arraybackslash}p{#1}}
\usepackage{dblfloatfix}
\usepackage{cuted}

\newcommand{\NCSO}{Na$_3$Co$_2$SbO$_6$ }
\newcommand{\NCTO}{Na$_2$Co$_2$TeO$_6$ }
\newcommand{\RuCl}{$\alpha$-RuCl$_3$ }
\newcommand{\BCAO}{BaCo$_2$(AsO$_4$)$_2$ }
\newcommand{\KCSO}{K$_2$Co(SeO$_3$)$_2$ }

\author{Naipeng Zhang}
\email{znaipeng@magnet.fsu.edu}
\affiliation{National High Magnetic Field Laboratory, Tallahassee, Florida 32310, USA}
\author{Nikolai Simonov}
\affiliation{School of Physics, Georgia Institute of Technology, Atlanta, Georgia 30332, USA}
\author{Mykhaylo Ozerov}
\affiliation{National High Magnetic Field Laboratory, Tallahassee, Florida 32310, USA}
\author{Sumedh Rathi}
\affiliation{School of Physics, Georgia Institute of Technology, Atlanta, Georgia 30332, USA}
\author{Nolan Heffner}
\affiliation{School of Physics, Georgia Institute of Technology, Atlanta, Georgia 30332, USA}
\author{Sara Huszar}
\affiliation{Department of Physics, Florida State University, Tallahassee, Florida 32306, USA}
\author{Long Chen}
\affiliation{Department of Physics and Astronomy, University of Tennessee, Knoxville, Tennessee 37996, USA}
\author{Haidong Zhou}
\affiliation{Department of Physics and Astronomy, University of Tennessee, Knoxville, Tennessee 37996, USA}
\author{Guangxin Ni}
\affiliation{Department of Physics, Florida State University, Tallahassee, Florida 32306, USA}
\alsoaffiliation{National High Magnetic Field Laboratory, Tallahassee, Florida 32310, USA}
\author{Chaebin Kim}
\affiliation{School of Physics, Georgia Institute of Technology, Atlanta, Georgia 30332, USA}
\author{Martin Mourigal}
\affiliation{School of Physics, Georgia Institute of Technology, Atlanta, Georgia 30332, USA}
\author{Stephen M. Winter}
\affiliation{Department of Physics and Center for Functional Materials, Wake Forest University, Winston-Salem, North Carolina 27109, USA}
\author{Zhigang Jiang}
\email{zhigang.jiang@physics.gatech.edu}
\affiliation{School of Physics, Georgia Institute of Technology, Atlanta, Georgia 30332, USA}
\author{Dmitry Smirnov}
\email{smirnov@magnet.fsu.edu}
\affiliation{National High Magnetic Field Laboratory, Tallahassee, Florida 32310, USA}

\title[An \textsf{achemso} demo]
  {Revealing Intrinsic Anisotropy of Collective Magnetic Excitations in Twinned Crystals of a Kitaev-Heisenberg Quantum Magnet}

\begin{document}

\let\oldmaketitle\maketitle
\let\maketitle\relax

\maketitle


\noindent
\textbf{Quantum magnets with competing interactions often emerge from delicate balances among microscopic parameters, making it essential to disentangle intrinsic spin dynamics from extrinsic disorder effects. Here, we introduce a multimodal optical approach combining magneto-infrared spectroscopy with domain-resolved micro-Raman spectroscopy at high magnetic fields to reconstruct the intrinsic magnetic excitation spectrum of twinned crystals of the Kitaev-Heisenberg quantum magnet Na$_3$Co$_2$SbO$_6$. Far-infrared spectroscopy reveals multiple field-tunable magnetic excitations, but the intrinsic response is obscured by replica features arising from twin domains. By correlating magneto-infrared and domain-resolved Raman spectra, we isolate the single-domain magnon response and uncover a pronounced twofold in-plane magnon anisotropy. This anisotropy far exceeds that expected from the measured in-plane g-factor anisotropy and is instead dominated by anisotropic bond-dependent exchange interactions. By unifying high-field, high-resolution and spatially selective optical probes, our work establishes a broadly applicable framework for revealing intrinsic spin dynamics and constraining the spin Hamiltonian in multidomain quantum magnets.}


\vspace{0.6em}






Magnons are the elementary low-energy excitations of ordered spin systems in crystals \cite{Balents2010a,Winter2017,Broholm2020}. Their spectra encode the underlying spin Hamiltonian and provide direct insight into magnetic ground states and phase transitions \cite{Winter2016,Takagi2019,Maksimov2020}. Magnon dispersions are typically measured by inelastic neutron scattering (INS), while zone-center modes can be resolved with high precision using optical probes such as terahertz and infrared (THz/IR) spectroscopy, providing essential reference points for modeling \cite{Jones2014,Zhang2018,Halloran2022,Na2024}. Unlike INS, which generally requires large sample volumes, optical techniques can be applied to substantially smaller single crystals \cite{Yankova2012,Kumawat2024}. In applied magnetic fields, the very high energy resolution of optical magneto-spectroscopy further enables accurate determination of the $g$-factors, imposing stringent constraints on spin models and their field evolution \cite{Sahasrabudhe2020, Wulferding2020,Wang2024,Zhang2023}.

However, optical spectroscopy is also highly sensitive to crystallographic disorder in quantum magnets. For example, far-infrared magneto-spectroscopy (FIRMS) measurements on the Ising-Heisenberg antiferromagnet \KCSO have shown that even minute amount of spin vacancies can generate sharp satellite lines adjacent to the main magnon modes, with intensities directly reflecting the vacancy density \cite{Kim2025}. In the Kitaev quantum spin liquid (QSL) candidate \NCTO,vacancies in the non-magnetic Na layers significantly modify the nearest-neighbor exchange interactions and local $g$-tensors of Co$^{2+}$ ions, leading to a substantial distribution of excitation energies and a broadened magnon dispersion \cite{Xiang2023}. Similarly, in another Kitaev QSL candidate \RuCl, interlayer stacking disorder gives rise to additional low-energy modes in THz spectroscopy, attributed to rigid-plane shear and breathing modes of molecular layers \cite{Reschke2019}. These examples underscore the importance of separating disorder-induced spectral features from intrinsic collective magnetic excitations in optical studies of quantum magnets.

Twinning represents another common form of crystallographic disorder, producing multidomain structures \cite{Cao2016, li2022giant, Breitner2023}. 
Because the beam size of THz/IR radiation or neutron probe volume generally exceed the characteristic domain size by orders of magnitude, FIRMS and INS measurements effectively average over multiple domains, producing overlapping spectral contributions that may mask the intrinsic spin dynamics of the system \cite{Wu2018,Lu2018}. Unlike FIRMS or INS, Raman magneto-spectroscopy can provide a direct probe of spin correlations \cite{Fleury1968,Knolle2014} with micrometer-scale spatial resolution \cite{Faugeras2011, Wulferding2020}.

In this Letter, we choose the Kitaev QSL candidate \NCSO \cite{Liu2020,Songvilay2020,Kim2022,Sanders2022,Liu-Kee2023}, in which crystal twinning is well documented \cite{li2022giant,Gu2024}, as a model system. 
We combine high-field FIRMS with micro-Raman magneto-spectroscopy to address the challenges posted by crystal twinning.
Using FIRMS, we observe a rich spectrum of magnetic excitations, including three distinct sets of excitations in the spin-polarized (SP) phase. By employing angle-resolved polarized Raman spectroscopy (ARPRS), we isolate the magnon modes associated with a single-crystal domain. Comparing the modes under different in-plane magnetic field directions further uncovers pronounced in-plane magnetic anisotropy. Leveraging high magnetic fields, we accurately determine the anisotropic in-plane $g$-factors and provide constraints on the anisotropic exchange parameters.

\begin{figure}[t]
\includegraphics[width=1\linewidth]{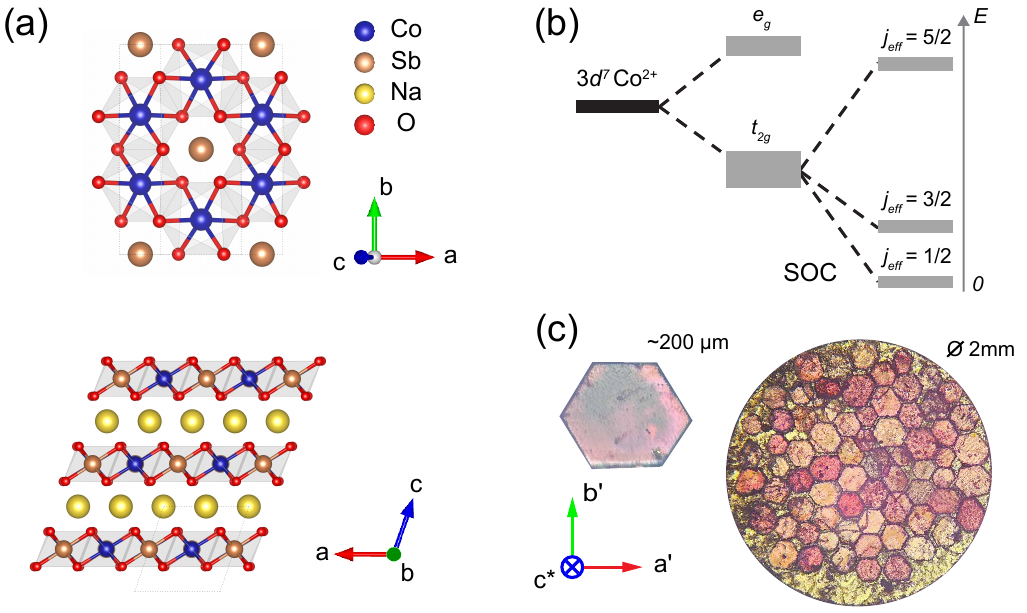}
\caption{Crystal structure and crystal-field levels in Na$_3$Co$_2$SbO$_6$.
(a) Crystal structure of \NCSO projected onto the $ab$ and $ac$ planes. The honeycomb basal-plane layers are monoclinically stacked along the $c$-axis with an offset of $-1/3$ of the lattice constant. 
(b) Splitting of Co$^{2+}$ 3$d^7$ states under the crystal field and SOC, resulting in $j_{\rm eff}=1/2$ ground state.
(c) Mosaic of oriented  \NCSO crystals assembled for FIRMS measurements. Inset: an example of an individual crystal used for micro-Raman spectroscopy measurements. Laboratory coordinate system $(a', b', c^*)$ defines the orientation of the mosaic with respect to the direction of the external magnetic field $B$.
}
\label{fig1}
\end{figure}

\begin{figure*}[t]
\begin{center}
    \includegraphics[width=\linewidth]{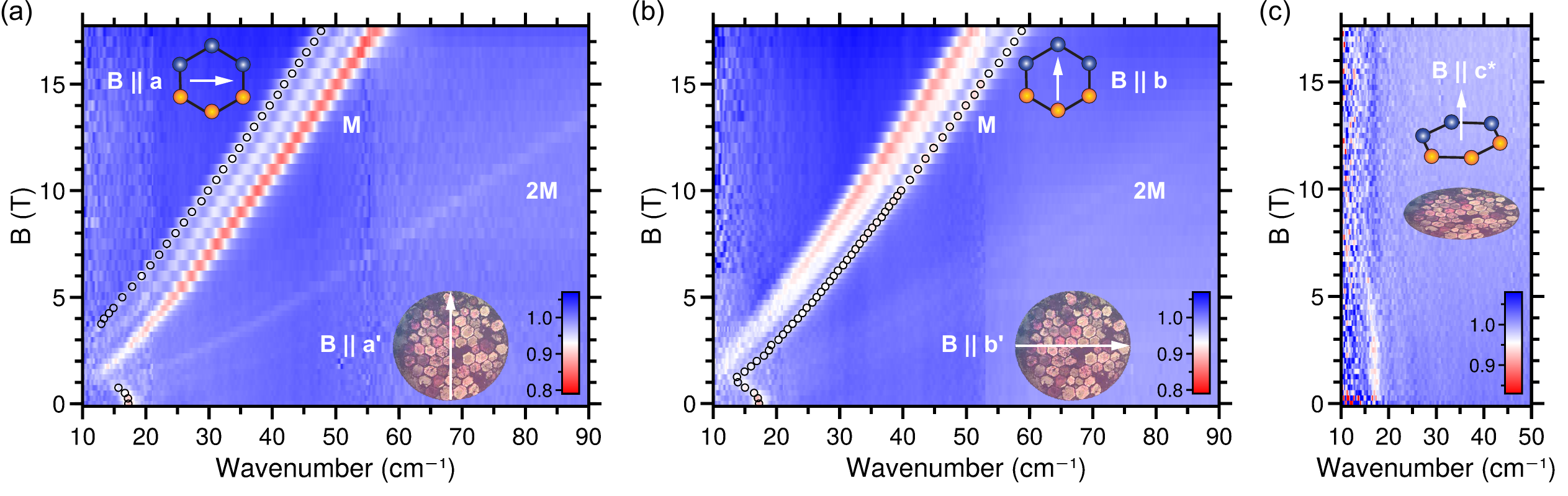}
    \caption{Magnetic excitations in a \NCSO mosaic revealed by FIRMS.
    Normalized FIRMS spectra measured at $T=5$~K on a mosaic of aligned \NCSO crystals, with the magnetic field nominally applied along the crystal edges $B \parallel a'$ (a), perpendicular to the edges $B \parallel b'$ (b), and  $B \parallel c^*$ (c). Open symbols indicate the peak positions of the one-magnon branches (M) corresponding to the “true” $B \parallel a$ and $B \parallel b$ geometries (illustrated in the inset), as determined by comparison with micro-Raman spectra measured on \NCSO single domains (Figure~\ref{fig3}).
    }
\label{fig2}
\end{center}
\end{figure*}

\NCSO crystallizes in a monoclinic layered structure with space group $C2/m$ \cite{Viciu2007}. Within the $ab$ plane, the Co$^{2+}$ ions, embedded in edge-sharing CoO$_6$ octahedra, form a two-dimensional honeycomb lattice (Figure~\ref{fig1}a). The honeycomb layers are stacked along the $c$-axis with a monoclinic offset of $-1/3$ of the lattice constant \textbf{a}, which breaks the $C_3$ rotational symmetry and retains only a $C_2$ in-plane rotational symmetry. Thus, the crystal structure of \NCSO closely resembles that of the monoclinic phase of \RuCl \cite{johnson2015, Cao2016, Nasir2026}, but differs from other cobaltates, such as \NCTO and \BCAO, where the in-plane $C_3$ symmetry is preserved \cite{Xiao2019, zhong2020}. The interplay of octahedral crystal field and spin-orbit coupling (SOC) yields an effective pseudospin-1/2 ground state (Figure~\ref{fig1}b), allowing for bond-dependent Kitaev interactions \cite{Liu2018, sano2018kitaev, Liu2020, liu2021towards, winter2022magnetic, Kim2022}.

\NCSO crystals were grown by the flux method \cite{yan2019}.  The crystals are plate-like hexagons with typical sizes up to 300 $\upmu$m. Due to the small orthorhombic distortion, twinned, multidomain crystals readily form, whereas single-domain crystals are relatively rare. For FIRMS measurements, we assembled a millimeter-sized mosaic of 68 individual crystals, both twinned and twin-free, aligned along the crystal edges, as shown in Figure~\ref{fig1}c. Here, we introduce an  orthogonal laboratory coordinate system $(a', b', c^*)$ to define the orientation of the mosaic with respect to the direction of an external magnetic field $B$. Micro-Raman magneto-spectroscopy measurements were performed on single domains.
Additional experimental details are provided in the Supporting Information.

\begin{figure*}[t!]
\begin{center}
    \includegraphics[width=0.8 \linewidth]{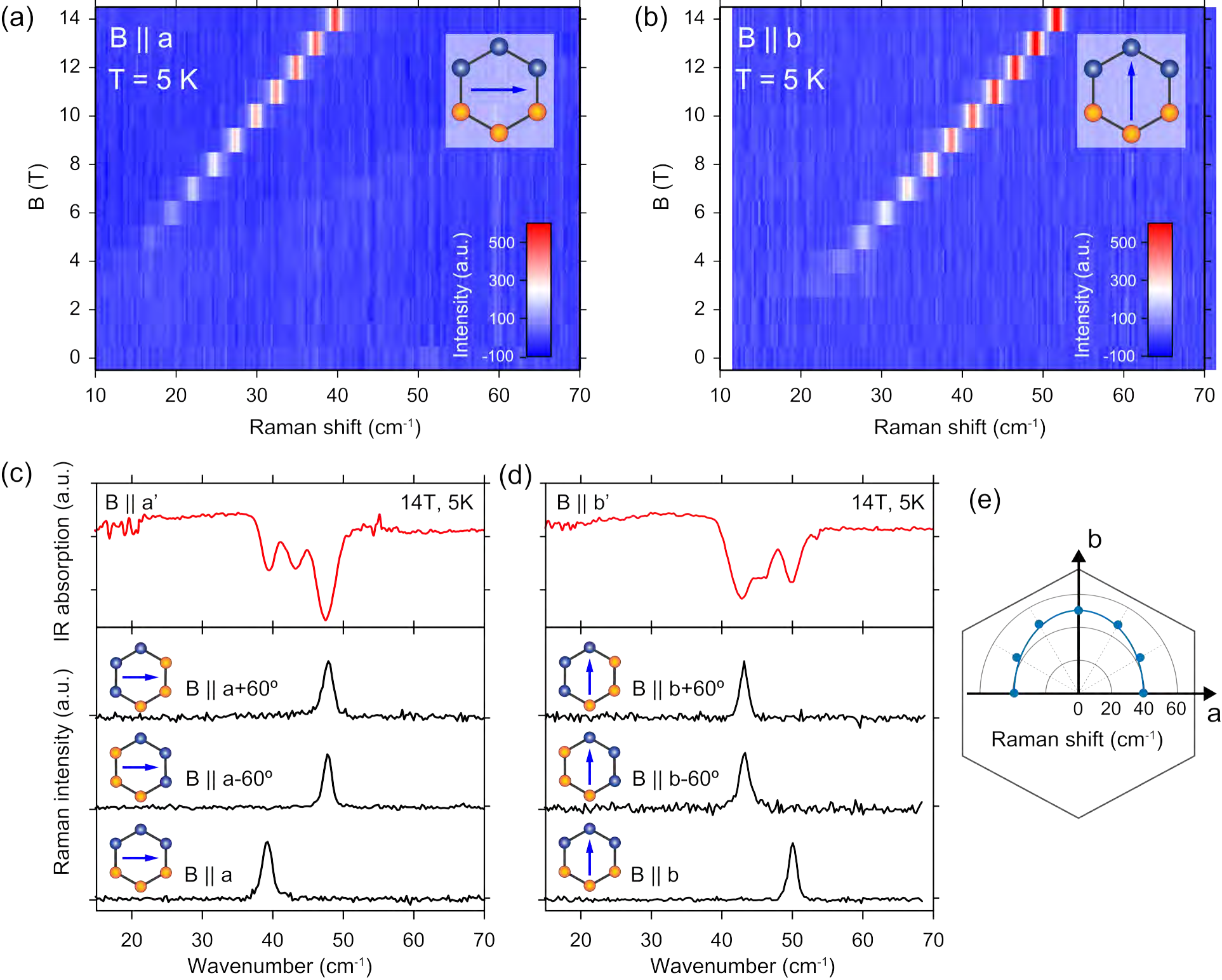}
    \caption{Deciphering the triple-peak structure in the FIRMS spectra through comparison with micro-Raman magneto-spectroscopy. (a,b) Raman spectra measured at $T=5$~K on a single domain of \NCSO for $B \parallel a$ and $B \parallel b$, respectively. (c,d) Top: FIRMS absorption spectra measured on the \NCSO crystal mosaic at $T=5$~K and $B=14$~T for $B \parallel a'$ and $B \parallel b'$, respectively. Bottom: Raman spectra collected from a single domain for in-plane magnetic fields oriented along $B \parallel a$ and $B \parallel a\pm60^\circ$ (c) or $B \parallel b$ and $B \parallel b\pm60^\circ$ (d), as illustrated in the insets. (e) Evolution of the one-magnon (M) mode at $B=14$~T upon rotating the single domain with respect to the in-plane magnetic field in $30^\circ$ steps.
    }
\label{fig3}
\end{center}
\end{figure*}

Figure~\ref{fig2} shows color maps of the normalized far-IR magneto-transmission spectra measured at $T=5$~K and $B$ up to 17.5~T on a mosaic of aligned \NCSO crystals in the Voigt geometry for $B\parallel a'$ (\ref{fig2}a) and $B\parallel b'$ (\ref{fig2}b), and in the Faraday geometry for $B\parallel c^*$ (\ref{fig2}c). Red color represents strong absorption. At $B=0$, a well-defined absorption peak at 17~cm$^{-1}$, corresponding to the zero-field gap of 2.1 meV in the magnetic excitation spectrum of the ordered antiferromagnetic (AFM) phase, is observed in all geometries, i.e. for both in-plane and out-of-plane magnetic fields. Under an in-plane field, this peak red-shifts and fades away when approaching the magnetic phase transitions, indicating spin-gap closing at a critical field $B_c$$\sim$1--2~T, in agreement with recent magnetic susceptibility measurements \cite{li2022giant,mi2025precisely-a91}.
As the in-plane field is further increased and the system enters the SP phase, a well-resolved set of three closely spaced absorption lines, labeled M in Figures~\ref{fig2}a and \ref{fig2}b, emerges and evolves nearly linearly with $B$, with similar slopes. A much weaker replica-like triple-peak structure, labeled 2M, is observed at approximately twice the energy and with twice the slope (Figures~\ref{fig2}a and \ref{fig2}b). This behavior is characteristic of $\Delta S=\pm 1$ magnetic-dipole transitions to magnon modes in an ordered magnetic state, allowing us to identify the observed spectral features as spin-flip one-magnon (M) and two-magnon (2M) excitations. Interestingly, at high in-plane magnetic fields, the 2M branch merges into another weak, replica-like triple-peak structure that exhibits the same slope as the one-magnon branch, as shown in Figure~S3. We tentatively attribute these modes to a higher-energy magnon branch made optically active by the interlayer coupling, zone folding, or disorder \cite{Cenker2021,Xiang2023}.

For an ideal twin-free \NCSO crystal, the Voigt-geometry FIRMS spectra in the SP phase are expected to exhibit a single one-magnon absorption line \cite{Li_2025}. In contrast, our measurements reveal three distinct, nearly parallel one-magnon branches under in-plane magnetic fields (Figures~\ref{fig2}a and \ref{fig2}b). We attribute this multiplicity to crystallographic twinning. As noted above, \NCSO adopts a monoclinic $C2/m$ crystal structure, which commonly gives rise to structural domains rotated by $\pm 60^\circ$ relative to one another during crystal growth. To isolate the magnetic response of a single domain, we first use ARPRS (Figure~S5) to map the structural domains within a \NCSO crystal, and then perform micro-Raman magneto-spectroscopy at $T=5$~K under in-plane magnetic fields up to $14$~T.

Figure~\ref{fig3} summarizes our domain-resolved measurements. Figures~\ref{fig3}a and \ref{fig3}b show magneto-Raman spectra collected from an individual domain, where only a single one-magnon excitation is observed for each in-plane field direction. By acquiring spectra in 30$^\circ$ rotation increments, we effectively reproduce the macroscopic FIRMS response. Specifically, the outer two modes of the triple-peak FIRMS structure in the SP phase can be assigned to one-magnon modes from the three possible twin orientations: $B \parallel a$ and $B \parallel a\pm 60^\circ$ for $B \parallel a'$ (Figure~\ref{fig3}c), and $B \parallel b$ and $B \parallel b\pm 60^\circ$ for $B \parallel b'$ (Figure~\ref{fig3}d). The weaker mode in the middle likely arises from slight misalignment among crystals in the mosaic or from magnetic scattering at domain walls (Figure~S7). These results demonstrate that the triple-peak structure observed in the macroscopic FIRMS measurements originates from spatial averaging over twinned domains. Furthermore, the angular dependence of the one-magnon mode at $14$~T (Figure~\ref{fig3}e) reveals the twofold symmetry of a single domain, where $E_a=39.3$~cm$^{-1}$, $E_b=50.0$~cm$^{-1}$, and $(E_b-E_a)/\sqrt{E_aE_b}=24.1\%$. In contrast, the in-plane $g$-factor anisotropy is much smaller. It can be estimated from the high-field slope of the magnon mode, $g'\propto \delta E/ \delta B$ (for $B\geq 10$~T), resulting in $(g'_b-g'_a)/\sqrt{g'_ag'_b}=5.7 \%$, thus indicating the dominant role of anisotropic exchange interactions over a weaker anisotropic Zeeman contribution.  

\begin{figure*}[t!]
\begin{center}
    \includegraphics[width=\linewidth]{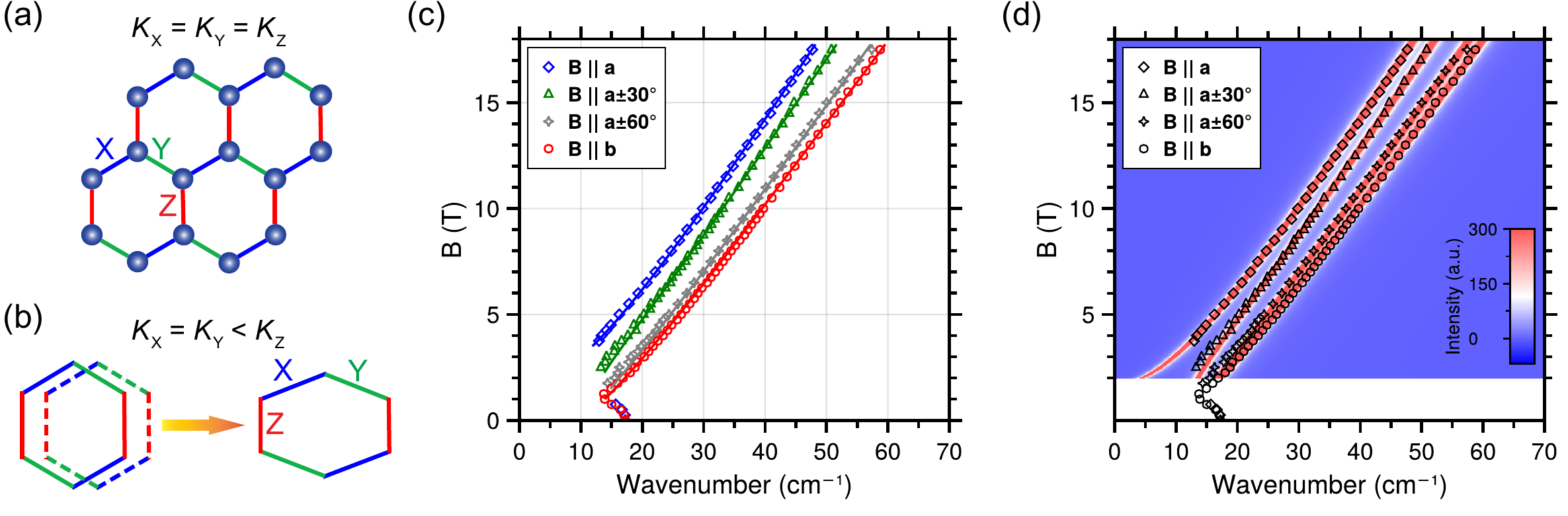}
    \caption{LSWT analysis of the anisotropic one-magnon mode in the SP phase of Na$_3$Co$_2$SbO$_6$. (a) Honeycomb lattice of Co$^{2+}$ ions, with blue, green, and red lines indicating the (X,Y,Z) bond-dependent nearest-neighbor exchange interactions, respectively. In the high-symmetry case, the Kitaev interactions satisfy $K_{\text{X}} = K_{\text{Y}} = K_{\text{Z}}$. 
    (b) Reduced symmetry arising from monoclinic stacking and in-plane distortion of the honeycomb lattice, resulting in anisotropic bond-dependent interactions, including $K_{\text{X}} = K_{\text{Y}} < K_{\text{Z}}$. (c) Experimental data (symbols) overlaid with CW fits of the one-magnon energies in the SP phase using eq~\ref{eqn:CW} for $\theta_B=0^\circ$, 30$^\circ$, 60$^\circ$, and 90$^\circ$, yielding $g_a = 4.75(2)$, $g_b = 5.32(2)$, $\Theta_{ba} = 2.73(5)$~K, and $\Theta_{bc^*} = 21.48(18)$~K.
    (d) LSWT results obtained using anisotropic exchange parameters constrained by the CW fits: $K_{\text{X}} = K_{\text{Y}} = -4.50$~meV, $K_{\text{Z}} = -4.00$~meV, $\Gamma_{\text{X}} = \Gamma_{\text{Y}} = 1.04$~meV, $\Gamma_{\text{Z}} = 1.50$~meV, and $\Gamma' = 0.56$~meV. Here, 1~meV$=8.0655$~cm$^{-1}$$=11.6045$~K$\cdot k_{\mathrm B}$.}
\label{fig4}
\end{center}
\end{figure*}

Although magneto-Raman measurements identify the one-magnon mode in the SP phase, FIRMS resolves a much richer spectrum, encompassing magnetic excitations in the AFM phase and across the field-induced phase transition. Guided by the Raman results, we isolate the single-domain one-magnon branches in the FIRMS spectra for $B \parallel a$ (open symbols in Figure~\ref{fig2}a) and $B \parallel a\pm 60^\circ$, as well as $B\parallel b$ (open symbols in Figure~\ref{fig2}b) and $B \parallel b\pm 60^\circ$. Crystallographically, $B \parallel b\pm 60^\circ$ is equivalent to $B \parallel a\pm 30^\circ$. The extracted mode energies, summarized in Figures~\ref{fig4}c and \ref{fig4}d, corroborate the pronounced in-plane anisotropy in the SP phase revealed by Raman measurements. In Figure~S8, we compare the observed energy anisotropy with the anisotropic Zeeman contribution, further confirming the dominant role of anisotropic exchange interactions, especially at low magnetic fields.

To model the one-magnon mode in the SP phase of Na$_3$Co$_2$SbO$_6$, we perform linear spin-wave theory (LSWT) calculations based on a generalized Kitaev-Heisenberg-$\Gamma$ Hamiltonian \cite{Rau2014,Rau2016}. Figures~\ref{fig4}a and \ref{fig4}b illustrate how the reduced symmetry arising from monoclinic stacking modifies the bond-dependent exchange interactions, such as Kitaev interactions $K_{\alpha}$ (where $\alpha=\text{X,Y,Z}$). The LSWT calculations indicate that a fully SP phase is reached for in-plane magnetic fields $B> 4$~T. In this regime, we 
first extend the phenomenological
Curie-Weiss (CW) relations of Ref.~\cite{Li_2025} to constrain the model parameters. For $B \parallel a \pm \theta_B$, the one-magnon energy reads
\begin{equation}
\begin{aligned}
&E_{\text{CW}}(\theta_B,\phi)
=
\sqrt{
E_\mathrm{Z}
+2k_{\mathrm B}
(\Theta_a-\Theta_b)\cos(2\phi)
}
\\[4pt]
&\times
\sqrt{
E_\mathrm{Z}
+2k_{\mathrm B}
\left(
\Theta_a\cos^2\phi
+\Theta_b\sin^2\phi
-\Theta_{c^*}
\right)
}.
\end{aligned}
\label{eqn:CW}
\end{equation}
where $E_\mathrm{Z} = \mu_{\mathrm{B}} B \left(g_a \cos \theta_B \cos\phi + g_b \sin \theta_B \sin \phi\right)$ is the Zeeman energy, $\theta_B$ is the angle of the magnetic field relative to $\mathbf{a}$, $\phi$ is the spin-polarization angle relative to $\mathbf{a}$, $\mu_{\mathrm B}$ is the Bohr magneton, $k_{\mathrm B}$ is the Boltzmann constant, and $\Theta_{a,b,c^*}$ are the CW temperatures along the corresponding crystallographic directions. The effects of anisotropic exchange interactions are encoded in the CW temperatures, as detailed in the Supporting Information. The angle $\phi$ is treated as a variational parameter and, for each magnetic field and trial parameter set, is determined self-consistently by minimizing the SP-state energy. The resulting $\phi$ is then used to evaluate the CW magnon energy (eq~\ref{eqn:CW}), and this procedure is iterated during the fitting with experimental data until convergence. Figure~\ref{fig4}c shows the results of the nonlinear least squares fit to the data for $B>4$~T for four field orientations, $\theta_B=0^\circ$, 30$^\circ$, 60$^\circ$, and 90$^\circ$, yielding $g_a = 4.75(2)$, $g_b = 5.32(2)$, $\Theta_{ba}\equiv\Theta_b-\Theta_a = 2.73(5)$~K, and $\Theta_{bc^*}\equiv\Theta_b-\Theta_{c^*} = 21.48(18)$~K. It may be noted that the extracted $g$-factors are significantly smaller than the $g_a$ and $g_b$ values deduced from the low-field magneto-THz data reported in Ref.~\cite{Li_2025}. As shown in the Supporting Information, the field dependence of the magnon mode remains non-linear over a large range of magnetic fields well above the critical field due to substantial contribution from the $B^{-1}$ and $B^{-2}$ terms. 

Next, we examine how anisotropic exchange interactions in \NCSO give rise to the large anisotropy observed in the magnon energies. Guided by prior understanding of the exchange parameters in honeycomb cobaltates \cite{Xiao2019,winter2022magnetic, Lee2025} and constrained by the CW fitting results, our LSWT calculations reveal that the one-magnon energy splitting is highly sensitive to bond-dependent exchange interactions, particularly $K$ and the symmetric off-diagonal interactions $\Gamma$ and $\Gamma'$. The data above the critical field ($B>2$~T) are quantitatively captured using $K_{\text{X}} = K_{\text{Y}} = -4.50$~meV, $K_{\text{Z}} = -4.00$~meV, $\Gamma_{\text{X}} = \Gamma_{\text{Y}} = 1.04$~meV, $\Gamma_{\text{Z}} = 1.50$~meV, $\Gamma' = 0.56$~meV, $g_a=4.75$, and $g_b=5.32$. Because $C_3$ symmetry breaking in $\Gamma$ and $\Gamma^\prime$ produces compensating effects, we retain an isotropic $\Gamma'$ and introduce the exchange anisotropy only through $\Gamma$. The calculated LSWT intensity map is overlaid with the experimental data in Figure~\ref{fig4}d, showing excellent agreement. Since LSWT does not precisely reproduce the critical field of the phase transition, we do not include the calculated spectra below 2~T. Further details of the model calculation are provided in Section VII of the Supporting Information. While the field evolution of the $\mathbf q = 0$ magnons is not sufficient to determine all exchange parameters independently, the present modeling results confirm that the one-magnon energy splitting in mosaic measurements arises from $C_3$ symmetry breaking.

In conclusion, we have combined high-field FIRMS and micro-Raman magneto-spectroscopy to disentangle the collective magnetic excitations of twinned Na$_3$Co$_2$SbO$_6$. FIRMS reveals a rich field-dependent excitation spectrum across the AFM-to-SP transition, including one-magnon, two-magnon, and higher-energy magnon branches. By comparing the macroscopic FIRMS response with domain-resolved micro-Raman measurements, we show that the prominent triple-peak structure in the SP phase arises primarily from spatial averaging over crystallographic twin domains. After isolating the single-domain response, we uncover a pronounced twofold in-plane anisotropy of the one-magnon energy that far exceeds the anisotropy expected from the measured $g$-factors. CW-constrained LSWT calculations further demonstrate that this anisotropy is governed by bond-dependent exchange interactions, particularly the Kitaev and symmetric off-diagonal terms. Our results establish combined magneto-infrared and Raman spectroscopy as a powerful route for resolving intrinsic magnetic excitations in multidomain quantum magnets and provide direct constraints on the anisotropic spin Hamiltonian of Na$_3$Co$_2$SbO$_6$.


\section{Supporting Information}
The Supporting Information is available free of charge at
https://pubs.acs.org/doi/... for experimental details, extended data (FIRMS, magneto-Raman), and LSWT calculations, which also cite Refs.~\cite{verble1970lattice,ponosov2024raman,xu2021,blundell2001magnetism,Sunny2025}.


\section{Notes}
The authors declare no competing financial interest.

\section{Acknowledgments}


This work was primarily supported by the DOE Basic Energy Sciences program (Grant No. DE-FG02-07ER46451). The crystal growth at UTK (L.C. and H.Z.) was supported by the Air Force Office of Scientific Research (Grant No. FA9550-23-1-0502). C.K. and M.M. (LSWT modeling) acknowledge support from the DOE BES program (Grant No. DE-SC0018660), and S.M.W. acknowledges support from the NSF (Grant No. DMR-2338704). N.G.X. acknowledges support from the DOE BES program (Grant No. DE-SC0022022) and from the NSF (Grant No. DMR-2145074). The FIRMS measurements were performed at NHMFL, which is supported by the NSF Cooperative Agreement (No. DMR-2128556) and the State of Florida. Theoretical calculations were supported in part through research cyber infrastructure resources and services provided by the Partnership for an Advanced Computing Environment (PACE) at Georgia Tech and the Center for Functional Materials at Wake Forest.

\bibliography{Main_ref}



\end{document}